\documentclass[draftcls,12pt,onecolumn]{IEEEtran}
\usepackage[utf8]{inputenc}
\usepackage{cite}
\usepackage{graphicx}
\usepackage{amssymb}
\usepackage{amsmath}

\ifCLASSINFOpdf

\else

\fi

\begin{document}

\title{Error Rate Analysis of GF($q$) Network Coded Detect-and-Forward Wireless Relay Networks Using Equivalent Relay Channel Models}

\author{Ilg{\i}n~{\c{S}}afak,~\IEEEmembership{Student~Member,~IEEE,}
        Emre~Akta{\c{s}},~\IEEEmembership{Member,~IEEE,}
        and~Ali~{\"O}zg{\"u}r~Y{\i}lmaz,~\IEEEmembership{Member,~IEEE}% <-this % stops a space
\thanks{I. {\c{S}}afak and E. Akta{\c{s}} are with the Department
of Electrical and Electronics Engineering, Hacettepe University, Beytepe Campus, Ankara 06800, Turkey, e-mail: ilgins@hacettepe.edu.tr, aktas@ee.hacettepe.edu.tr.}% <-this % stops a space
\thanks{A. {\"O}. Y{\i}lmaz is with the Department of Electrical and Electronics Engineering, Middle East Technical University, Ankara 06531, Turkey, e-mail: aoyilmaz@metu.edu.tr.}% <-this % stops a space
%\thanks{Manuscript received April 19, 2005; revised January 11, 2007.}
}

% make the title area
\maketitle

\begin{abstract}
%\boldmath
This paper investigates simple means of analyzing the error rate performance of a general $q$-ary Galois Field network coded detect-and-forward cooperative relay network with known relay error statistics at the destination. Equivalent relay channels are used in obtaining an approximate error rate of the relay network, from which the diversity order is found. Error rate analyses using equivalent relay channel models are shown to be closely matched with simulation results. Using the equivalent relay channels, low complexity receivers are developed whose performances are close to that of the optimal maximum likelihood receiver.
\end{abstract}


\section{Introduction}

\IEEEPARstart{C}{ooperative} communication is effective in extending the coverage area and increasing the quality of service of relay networks by means of diversity \cite{Proakis2000,Liu2008}. Amplify-and-forward (AF) and decode-and-forward (DF) are two commonly used techniques for relaying. In AF relaying, the signal received at the relay is transmitted after being amplified by a factor \cite{Liu2008,Hasna2004}. In DF, the received signal is decoded then re-encoded and modulated at the relay before re-transmission. AF needs redesign of many transceivers while DF is computationally complex for the relay node \cite{Liu2008,Wang2007,Li2010}. The detect-and-forward (DetF) technique, where the received symbol is detected and then modulated without decoding and re-encoding, is a low-complexity alternative to DF \cite{Benjillali2010,Chen2006,Gao2011,Yuan2010,Chu2008}. In this paper, the DetF method is considered.

The simplest method of transmitting a message in a cooperative communication network is routing, where the data received at a node are simply forwarded to another \cite{Liu2008}. Alternatively, in network coding (NC), multiple data received at a node can be combined and transmitted simultaneously. Hence, NC can be used to increase the throughput of the system and reduce bandwidth and transmission energy consumption \cite{Ahlswede2000}.

Non-binary NC is a simple and efficient means of achieving improved diversity gain in wireless relay networks compared to binary NC.
In \cite{Lei}, a $q$-ary Galois field (GF($q$)) NC multi-user cooperative wireless relay network is considered. The users act as each other's relay nodes and use GF($q$) NC and decode-and-forward (DF) techniques. The use of non-binary NC was shown to lead to higher diversity orders compared to binary NC \cite{Lei}.
In \cite{Xiao2009b}, a maximum diversity order achieving non-binary NC scheme, namely dynamic NC (DNC), is devised for cooperative wireless DF networks in a block fading environment. The network topology is assumed to be dynamic and the network code is chosen to be deterministic until the topology changes. Perfect error detection is assumed at the relays; erroneous blocks at the relays are discarded, resulting in erasure channels.
In \cite{Xiao2009a}, a GF($q$) NC DF, multiple-user, multiple-relay cooperative wireless network with slow fading channels is considered. The asymptotic performance of the relay network is investigated using the diversity order expression, which is derived in the high SNR regime. It is shown that a maximum diversity order of $N+1$ can be obtained for a system of $N$ users.
In \cite{Rebelatto2010}, the DNC method in \cite{Xiao2009b} is generalized for a cooperative $M$-user single-destination network where multiple packets are allowed to be sent by the relays. An outage probability analysis is given, where the proposed scheme, generalized DNC (GDNC), is shown to display a better trade-off between rate and diversity.
In \cite{Rebelatto2011}, an adaptive distributed network-channel coding scheme based on GDNC in \cite{Rebelatto2010} is proposed, where $M$-users cooperatively communicate with a common destination in a Rayleigh fading environment. Feedback is used at the destination to increase the average code rate without reducing the diversity order of the system. Average error rate and diversity gain expressions are obtained for high SNR values.

Availability of channel state information (CSI) or SNR values at the relay and/or at the destination improves the performance of relay networks significantly. If CSI is available at an AF relay of a source-relay-destination $(S-R-D)$ link, the amplification factor may be chosen to compensate for the instantaneous channel gain of the $S-R$ link \cite{Hasna2004}. If SNR values of the $S-R$ link are available at the destination in DetF relay networks, the error propagation effects can be mitigated by taking the relay detection error statistics into account at the destination \cite{Xiao2009}.

The closed-form error rate of a GF(2) NC DetF relay network with known relay error statistics is obtained in \cite{Iezzi2011,Iezzi2012}, where it is shown that there is a trade-off between diversity, coding gain and throughput efficiency in \cite{Iezzi2012}. The error performance analyses in \cite{Xiao2009,Iezzi2011,Iezzi2012,Li2010} are derivable when GF(2) NC and BPSK modulation techniques are employed, however, the analyses are highly complex for higher Galois fields and modulation orders. Various analytically tractable approximations have been utilized in the literature in reducing the mathematical complexity of the error rate analysis. In \cite{Benjillali2010}, the max-log approximation is used in the performance analysis of a coded cooperative, two-hop DetF relay network with multiple relays. The approximation exhibits accurate error performance for the relay network. However, the approximation in \cite{Benjillali2010} does not consider network-coded relay networks. Another approximation technique is the equivalent relay channel method, which approximates a two-hop relay channel with a single-hop channel \cite{Gao2011,Yuan2010,Chu2008,Wang2007,Li2010,Kim2011}. Using an equivalent relay channel model, a maximum diversity achieving receiver is derived in \cite{Wang2007}, and error rate expressions are obtained for it, where NC is not applied.
The equivalent relay channel model developed for multi-hop DF relay networks without NC in \cite{Wang2007}, is shown to be valid also for GF(2) NC DF two-way relay channels in \cite{Li2010}, where closed error rate expressions are obtained.
In \cite{Nasri2010}, a GF($q$) NC DetF Rayleigh fading cooperative wireless relay network with multiple sources, single relay and joint destination is considered.
In \cite{Kim2011}, an equivalent relay channel method called propagation error modeling (PEM) is devised for a low-complexity detector for GF($q$) and complex field network coded (CFNC) DF relay networks. In PEM, the detection errors that occur at the relays are modeled as virtual noise at the destination in order to overcome the error propagation effects in DF systems. The PEM method, which displays close performance to the relay network, can be used in obtaining the diversity order of the relay network. However, this method does not readily provide the error rate.

To the best of our knowledge, simple means of obtaining the diversity order and error rate for ML detection of general GF($q$) NC DetF relay networks which use relay error statistics at the destination is lacking. Low-complexity alternatives to the optimal receiver for such a relay network are also not available. The aim of this paper is to address these problems.

In this contribution, the error rate analysis methods of \cite{Xiao2009,Wang2007} are generalized to GF($q$) NC DetF relay networks. Two-hop relay channels are approximated using equivalent relay channel models,  where the equivalent relay channel model in \cite{Wang2007} is adapted to general GF($q$) NC DetF relay networks. The equivalent relay models are then used in finding pairwise error probabilities (PEPs), from which the diversity order of ML detection and an approximation to the union bound are obtained. The optimal receiver's error performance is compared with that of the receiver which performs ML detection under the assumption of an equivalent relay channel. It is observed that the equivalent channel models successfully approximate the relay network. The equivalent relay channel is observed to be a suitable tool both for the error rate analysis of the GF($q$) NC system and also for building low-complexity sub-optimal NC receivers.

The organization of this paper is as follows. In Section II, the considered system model, data transmission technique, equivalent relay channel model and receiver structures are given and the ML rule is derived. In Section III, error rate analyses of relay networks with error-free and error-prone $S-R$ channels are presented; an approximate union bound for error rates is obtained using equivalent relay channel models. In Section IV, the ML rule and union bound symbol error rate (SER) performances of the equivalent relay networks are obtained and compared to the relay network's optimum receiver performance. The results are summarized in Section V.

\section{GF($q$) Network Coded Detect-and-Forward Relay Network}

\subsection{System Model and Data Transmission}
\begin{figure}[!t]
\centering
\includegraphics[width=3in]{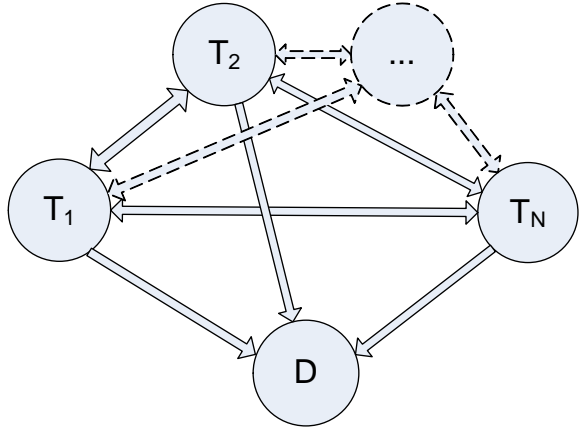}
\caption{$N$-user relay network}
\label{relay}
\end{figure}

As a general model, the relay network model in Fig. \ref{relay} is considered. $M$-ary phase shift keying (PSK) and time division multiplexing (TDM) techniques are used in the transmission of the GF($q$) network coded symbols. All channels are assumed to be independent and undergo Nakagami-$m$ fading. The users act as each other's relay nodes, where the DetF method is used in relaying. Imperfect detection at the relays is taken into account, and the detection error statistics observed at the relays are assumed to be available at the destination for ML detection. The users and destination are assumed to have SNR knowledge of the incoming data's channel.

Data transmission of $N$ user terminals is completed in $K$ time-slots. Each user terminal $T_{n}$, $1 \leq n \leq N$, broadcasts its modulated data in the $n^{th}$ time-slot to the other terminals and the destination $D$, where the data received at the terminals are detected. The destination D is an arbitrary terminal which is taken to be distinct from the user terminals without loss of generality. After all users have completed transmitting their data, $K-N \:(K>N)$ users transmit modulated network coded  data to $D$ using a pre-determined NC rule, hence a total of $K$ signals are received at $D$. The received signals for $K$ time-slots at $D$ can be expressed as
\begin{equation}\label{y}
    \mathbf{y_{D}}=\mathbf{H_{D}}\,\mathbf{s} + \mathbf{n_{D}},
\end{equation}
where $\mathbf{y_{D}}$ is the $K \times 1$ sized vector consisting of $K$ signals received at $D$:
\begin{equation}\label{yD}
\begin{array}{cccccc}
\mathbf{y_{D}}=[y_{1D,sys} & \ldots & y_{ND,sys} & y_{1D,nc} & \ldots & y_{(K-N)D,nc}]^{T}.
\end{array}
\end{equation}
In (\ref{yD}), $y_{nD,sys}$ is the observation at D corresponding to the $n^{th}$ direct transmission, which is the systematic symbol from $T_{n}$; $y_{lD,nc}$ is the observation at D corresponding to the $l^{th}$ network coded data, $1 \leq l \leq K-N$; $\mathbf{H_{D}}$ is the $K \times K$ sized diagonal fading coefficient matrix observed at $D$; $\mathbf{n_{D}}$ is the $K \times 1$ sized noise vector at $D$ and $\mathbf{s}$ is the $K \times 1$ sized vector comprising the network coded modulated data transmitted from the terminals:
\begin{equation}\label{s}
    \mathbf{s}=\varphi\left(\mathbf{G^{T}}\mathbf{u}+\mathbf{e_{R}}\right).
\end{equation}
Here, $\varphi(\cdot)$ denotes the constellation mapping and $\mathbf{G}$ is the generator matrix which represents how NC is performed in the relay network. For the considered NC operation, $\begin{array}{ccc}
\mathbf{G}=[\mathbf{I} & \vdots & \mathbf{P}]_{N \times K}
\end{array}$, where the $N \times N$ identity matrix $\mathbf{I}$ represents the systematic data (first $N$ symbol transmissions), and  the $(K-N) \times K$ parity  matrix $\mathbf{P}$ represents the remaining $K-N$ network coded symbols transmitted. The data vector sent by the user terminals is shown by $\begin{array}{ccc}
\mathbf{u}=[u_{1} & \ldots & u_{N}]
\end{array}
$, where $u_{n}$, $1 \leq n \leq N$, is the data of $T_{n}$;
$\mathbf{e_{R}}$ is the $K \times 1$ sized coded detection error vector at the relays. The elements of $\mathbf{G},\mathbf{u}$ and $\mathbf{e_{R}}$ are in GF($q$).

Let us define the set of all modulated GF($q$) network coded codewords $\chi$. When the relay network in Fig. \ref{relay} has error-free $S-R$ links, there are $q^N$ possible such codewords. Corresponding to each source symbol configuration is a codeword $\mathbf{X}(i)$, where $\mathbf{X}(i)$ is the $i^{th}$ element of the codeword set $\chi$:
\begin{equation}\label{X}
    \mathbf{X}(i)=\varphi(\mathbf{G^{T}}\mathbf{u})
    =\varphi([c_{1} \ldots c_{K}]),
\end{equation}
assuming the $i^{th}$ element of the set of all data vectors of $\mathbf{u}$ is transmitted. The aim of the destination is to decide which codeword was sent based on the observation $\mathbf{y_{D}}$ and the error statistics of the relays.

\subsection{Optimum Receiver}\label{section:optimum}
The \emph{optimum soft-receiver} performs ML detection based on the soft-decision observation vector in (\ref{y}), whereas the \emph{optimum hard-receiver} performs ML detection based on the hard-decision observation vector in (\ref{Z}):
\begin{equation}\label{Z}
\begin{aligned}
&\mathbf{y_{D}} =\begin{array}{cccccc}
[y_{1D,sys} & \ldots & y_{ND,sys} & y_{1D,nc} & \ldots & y_{(K-N)D,nc}]^{T}
\end{array}\\
&\overset{hard-decision}{\rightarrow}\mathbf{z_{D}}=\begin{array}{cccccc}
[z_{1D,sys} & \ldots & z_{ND,sys} & z_{1D,nc} & \ldots & z_{(K-N)D,nc}]^{T}
\end{array}.
\end{aligned}
\end{equation}
$y_{nD,sys}$ and  $y_{mD,nc}$ denote the $n^{th}$ systematic and $m^{th}$ network coded data, respectively; $z_{nD,sys}$ and $z_{mD,nc}$ are the hard-decision data obtained by individually detecting $y_{nD,sys}$ and $y_{mD,nc}$ respectively.
In relay networks with error-free $S-R$ channels, the \emph{soft-decision ML} rule of terminal $T_i$'s data $u_{i},\,1\leq i \leq N$, is given as:
\begin{equation}\label{soft_MLD1}
  u_{i}^{ML}= \arg\max_{u_{i}} p(\mathbf{y_{D}}|u_{i}).
\end{equation}
The likelihood function given above is expanded as
\begin{equation}\label{soft_MLD2}
\begin{aligned}
  p(\mathbf{y_{D}}|u_{i})  &= \sum_{\bf{u}^{i}}   p(\mathbf{y_{D}}|u_{i},\mathbf{u}^{i})p(\mathbf{u}^{i})\\
  &=\sum_{\bf{u}^{i}}  \prod_{i=1}^{N} p(y_{iD,sys}|u_{i})
  \prod_{j=1}^{K-N} p(y_{jD,nc}|u^{i},\mathbf{u}^i) p(\mathbf{u}^{i}),
\end{aligned}
\end{equation}
where $\mathbf{u}^{i}=\mathbf{u} \setminus u_{i}$ is the data vector excluding $u_{i}$. Assuming that $q$-ary symbols of all $N$ users are transmitted equally likely and independently, the likelihood function in (\ref{soft_MLD2}) becomes
\begin{equation}\label{soft_MLD3}
  p(\mathbf{y_{D}}|u_{i})
  = q^{-(N-1)} \sum_{\bf{u}^{i}}  \prod_{i=1}^{N} p(y_{iD,sys}|u_{i})
  \prod_{j=1}^{K-N} p(y_{jD,nc}|u^{i},\mathbf{u}^i).
\end{equation}
In relay networks with error-prone $S-R$ channels, when no relay error statistics are available at the destination, the soft-decision ML rule is equivalent to that used in the error-free case, given by (\ref{soft_MLD2}). However, if the relay error statistics are known at $D$, the likelihood function for the soft-decision ML rule is
\begin{equation}\label{sMLD_CSI}
 \begin{aligned}
  p(\mathbf{y_{D}}|u_{i})&= \sum_{\mathbf{e_{R}}} p(\mathbf{y_{D}}|u_{i},\mathbf{e_{R}})p(\mathbf{e_{R}})\\
  &= \sum_{\mathbf{e_{R}}}\sum_{\mathbf{u}^{i}}  \prod_{i=1}^{N} p(y_{iD,sys}|u_{i})
  \prod_{j=1}^{K-N} p(y_{jD,nc}|u^{i},\mathbf{u}^i,\mathbf{e_{R}}) p(\mathbf{u}^{i})p(\mathbf{e_{R}}),
 \end{aligned}
\end{equation}
where the fact that relay events are independent of other events is used. Assuming equally likely  and independently transmitted symbols, the optimal soft-decision ML rule which takes relay errors into account is
\begin{equation}\label{py_u}
\begin{aligned}
  p(\mathbf{y_{D}}|u_{i})
  &= q^{-(N-1)} \sum_{\mathbf{e_{R}}}\sum_{\mathbf{u}^{i}}  \prod_{i=1}^{N} p(y_{iD,sys}|u_{i})
  \prod_{j=1}^{K-N} p(y_{jD,nc}|u^{i},\mathbf{u}^i,\mathbf{e_{R}}) p(\mathbf{e_{R}}).
\end{aligned}
\end{equation}

The \emph{hard-decision ML rule} of terminal $u_{i},\,1\leq i \leq N$ is
\begin{equation}\label{hard_MLD}
  u_{i}^{ML}= \arg\max_{u_{i}} p(\mathbf{z_{D}}|u_{i}).
\end{equation}
Assuming $q$-ary symbols of all $N$ users are transmitted equally likely and independently,  the likelihood function given in (\ref{hard_MLD}) is expanded as
\begin{equation}\label{hard_MLD2}
\begin{aligned}
p(\mathbf{z_{D}}|u_{i})=q^{-(N-1)} \sum_{\mathbf{u}^{i}}  \prod_{i=1}^{N} p(z_{iD,sys}|u_{i})
  \prod_{j=1}^{K-N} p(z_{jD,nc}|u^{i},\mathbf{u}^i).
\end{aligned}
\end{equation}
Similar to the soft-decision case, the optimal hard-decision ML rule which takes relay errors into account is
\begin{equation}\label{hMLD_CSI}
  p(\mathbf{z_{D}}|u_{i})= q^{-(N-1)} \sum_{\mathbf{e_{R}}}\sum_{\mathbf{u}^{i}}  \prod_{i=1}^{N} p(z_{iD,sys}|u_{i})
  \prod_{j=1}^{K-N} p(z_{jD,nc}|u^{i},\mathbf{u}^i,\mathbf{e_{R}}) p(\mathbf{e_{R}}).
\end{equation}
Compared to (\ref{soft_MLD3}) and (\ref{hard_MLD2}), the likelihood functions in (\ref{py_u}) and (\ref{hMLD_CSI}) involve the averaging over $\mathbf{p(e_{R})}$, which increases the mathematical complexity of the soft- and hard-ML rules, respectively, hence increasing the computational complexity of the optimal soft- and hard-receivers.

\subsection{Equivalent Receiver}\label{section:equivalent}
The computational complexity of the optimal receiver can be decreased by using an equivalent receiver, where the multi-source relay channels in the relay network in Fig. \ref{relay} are approximated by equivalent relay channels as follows.

\begin{figure}[h]
\centering
\includegraphics[width=4in]{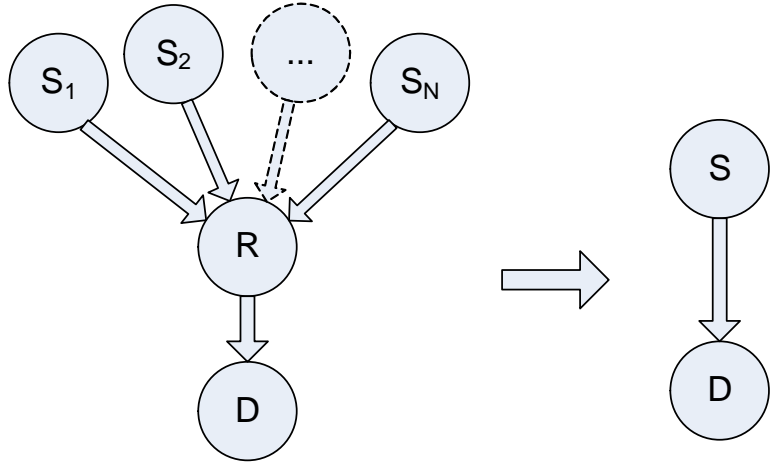}
\caption{Equivalent relay channel of a relay channel with $N$ sources}
\label{GFq_equivalent}
\end{figure}

Data received from the terminals $T_{\tilde{n}}, 1 \leq \tilde{n} \leq N, \tilde{n} \neq n,$  are detected and then GF($q$) network coded at the relay terminal $T_{n}$
\begin{equation}\label{uR}
    \hat{u}_{nR}=u_{nR}+e_{nR},
\end{equation}
where ${u}_{nR}$ is the error-free network coded data to be transmitted from $T_{n}$; $e_{nR}$ is the network coded detection error at $T_{n}$; $\hat{u}_{nR}, u_{nR}$ and $e_{nR}$ are in GF($q$).
The $m^{th}$ network coded data received at $D$ is
\begin{equation}\label{yD_nc}
    y_{mD,nc}=h_{nD,nc}s_{nR}+n_{mD,nc},
\end{equation}
where $1 \leq m \leq K-N$; $h_{nD,nc}$ is the fading coefficient of the $T_{n}-D$ channel over which the network coded data is sent; $s_{nR}=\varphi (\hat{u}_{nR})$ is the modulated network coded data sent from $T_{n}$ and $n_{mD,nc}$ is the noise term at the destination of the observed network coded data. The multi-source relay channel described above can be approximated by a single-hop $S-D$ channel as shown in Fig. \ref{GFq_equivalent}, where $T_{n}$ corresponds to the relay node $R$ and terminals $T_{\tilde{n}}$ represent the source nodes $S_{\tilde{n}}, \tilde{n} \neq n$.

Assuming soft-decision ML detection and known error statistics at $D$, the likelihood function of the channel over which the network coded data to be used is sent in (\ref{py_u}) as follows:
\begin{equation}\label{likelihood_eq}
  p(y_{mD,nc}|u_{nR})=\sum_{e_{nR}=1}^{q-1} p(y_{mD,nc}|\hat{u}_{nR},e_{nR}) p(e_{nR}).
\end{equation}
Dropping the $m's$ and $n's$, the likelihood function given in (\ref{likelihood_eq}) is approximated with the likelihood function of a single-hop channel:
\begin{equation}\label{Peq_sML}
 \begin{aligned}
  &p(y_{D,nc}|u_{R}) \simeq p_{eq}(y_{D,nc}|u_{R})=\frac{1}{ \pi \sigma^{2}_{eq}}\exp\left(-\frac{1}{\sigma^{2}_{eq}}\left|y_{D,nc}-h_{eq}\varphi(u_R)\right|^2\right),
\end{aligned}
\end{equation}
where $h_{eq}$ is the fading coefficient and $\sigma^{2}_{eq}$ is the noise variance of the equivalent relay channel.

When hard-decision ML decoding is performed, the hard-decision of the network coded data in (\ref{yD_nc}) is used.
The instantaneous end-to-end symbol error probability (SEP) of the relay channel is
 \begin{equation}\label{Peq_Nuser}
  \mathrm{Pr}({z}_{mD,nc} \neq u_{nR}|u_{nR})= \sum_{e_{nR}=1}^{q-1} p({z}_{mD,nc} \neq u_{nR}|\hat{u}_{nR}=u_{nR}+e_{nR})p(e_{nR}).
\end{equation}
Dropping the $m's$ and $n's$, the instantaneous SEP given in (\ref{Peq_Nuser}) is approximated by the instantaneous SEP of a single-hop channel which has an instantaneous SNR per symbol $\gamma_{eq}$:
\begin{equation}\label{Peq}
\begin{aligned}
    P_{eq}^{s}(\gamma_{eq}) &\simeq \mathrm{Pr}({z}_{D,nc} \neq u_{R}|u_{R})\\
    &= \sum_{e_{R}=1}^{q-1} p({z}_{D} \neq u_{R}|\hat{u}_{R}=u_{R}+e_{R})p(e_{R}),
\end {aligned}
\end{equation}
where $P_{eq}^{s}(\gamma_{eq})$ is determined by the modulation constellation.

\subsubsection*{Q-inverse Equivalent Receiver}
\begin{figure}[!t]
\centering
\includegraphics[width=5in]{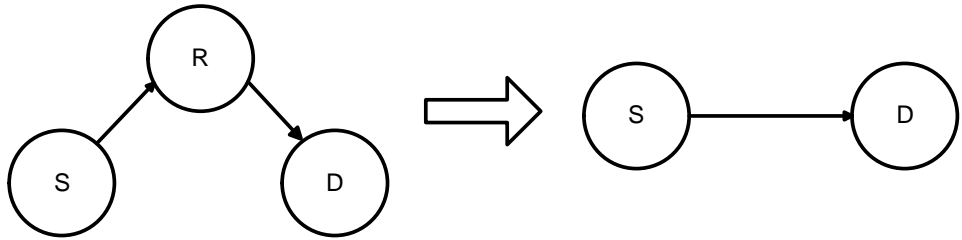}
\caption{Equivalent relay channel of a single-sourced relay channel}
\label{equivalent}
\end{figure}

In \cite{Wang2007}, an equivalent relay channel model is proposed that approximates a multi-hop, cooperative DF relay channel with a single-hop channel and the model is shown to be valid for any modulation constellation, assuming coherent modulation/demodulation. The equivalent relay channel takes the channel model of the $S-R$ and $R-D$ channels, which is exemplified by a flat Rayleigh fading, two-hop relay channel with coherent BPSK modulation/demodulation, as shown in Fig. \ref{equivalent}. The instantaneous bit-error probability (BEP) of the equivalent relay channel is found as \cite{Wang2007}
\begin{equation}\label{Peq_Qinv11}
    P^{b}_{Q^{-1}}=P^{b}_{SR}\left(1-P^{b}_{RD}\right)+P^{b}_{RD}\left(1-P^{b}_{SR}\right),
\end{equation}
where $P^{b}_{SR}$ and $P^{b}_{RD}$ denote the BEP's of the $S-R$ and $R-D$ links, respectively. Assuming coherent BPSK modulation/demodulation, the equivalent BEP expressed above is approximated with the BEP of a single-hop channel that has an instantaneous SNR per bit $\gamma^{b}_{Q^{-1}}$ \cite{Wang2007}:
\begin{equation}\label{Peq_Qinv12}
\begin{aligned}
    P_{Q^{-1}}^{b}& = Q\left(\sqrt{2\gamma^{b}_{SR}}\right)\left(1-Q\left(\sqrt{2\gamma^{b}_{RD}}\right)\right)+Q\left(\sqrt{2\gamma^{b}_{RD}}\right)\left(1-Q\left(\sqrt{2\gamma^{b}_{SR}}\right)\right)\\
    & = Q\left(\sqrt{2\gamma^{b}_{Q^{-1}}}\right),
\end{aligned}
\end{equation}
where $\gamma^{b}_{SR}$ and $\gamma^{b}_{RD}$ denote the instantaneous SNR's per bit of the $S-R$ and $R-D$ links, respectively.
From (\ref{Peq_Qinv12}), an instantaneous SNR value of the equivalent relay channel is found as \cite{Wang2007}
\begin{equation}\label{SNR_Qinv}
    \gamma_{Q^{-1}}=\frac{1}{2}\{Q^{-1}(P_{Q^{-1}}^{b})\}^{2}.
\end{equation}
Due to the use of the Gaussian Q-inverse function in (\ref{SNR_Qinv}), the equivalent relay channel model is referred to as the Q-inverse equivalent relay channel in the rest of this paper. Notice from (\ref{Peq_Qinv11}) and (\ref{Peq_Qinv12}) that, for a fading channel, the instantaneous SNRs $\gamma^{b}_{SR}$ and $\gamma^{b}_{RD}$ are random variables, thus the instantaneous SNR $\gamma^{b}_{Q^{-1}}$ is also a random variable. For the error rate analysis, the distribution of $\gamma^{b}_{Q^{-1}}$ is approximated by a conventional fading SNR distribution whose parameters are to be determined.

The Q-inverse equivalent relay channel model, originally developed for DF relay networks, is used in a GF(2) NC DF two-way relay channel in \cite{Li2010} and has the same BEP expression as in (\ref{Peq_Qinv12}) and SNR value in (\ref{SNR_Qinv}).

The novel part of this paper involves the adaptation of the Q-inverse equivalent relay channel model to GF($q$) NC, coherent $M$-PSK modulated, DetF, Nakagami-m fading relay channels with multiple sources. The proposed equivalent relay channel model is used in obtaining approximate error rate expressions for GF($q$) NC DetF, coherent $M$-PSK modulated, Nakagami-m fading relay networks. Soft- and hard-decision ML receivers employing the Q-inverse equivalent relay channel model, which display very close performance to the optimal receiver, are developed. Assuming hard-decision ML detection at $D$, the instantaneous SEP in (\ref{Peq_Nuser}) is approximated to the instantaneous SEP of a single channel \cite{Safak2013}
\begin{equation}\label{GFq_Peq}
\begin{aligned}
    P_{Q^{-1}}^{s}&= 2\,Q\left(\sqrt{2\gamma_{Q^{-1}}}\, \sin\left(\pi/M\right)\right)\\
    & \simeq \mathrm{Pr}({z}_{D,nc} \neq u_{R}|u_{R})
\end{aligned}
\end{equation}
for coherent $M$-ary PSK for large values of $M$ and SNR \cite{Proakis2000}. The instantaneous SNR per symbol of the Q-inverse equivalent relay channel is found using (\ref{GFq_Peq}) as  \cite{Safak2013}:
\begin{equation}\label{GFq_SNR}
\begin{aligned}
    \gamma_{Q^{-1}} &= \frac{1}{2\sin^{2}\left(\pi/M\right)}\left\{ Q^{-1}(0.5 P_{Q^{-1}}^{s})\right\}^{2}\\
    &=\frac{1}{2\sin^{2}\left(\pi/M\right)}\left\{ Q^{-1}\left(0.5 \sum_{e_{R}=1}^{q-1} p({z}_{D,nc} \neq u_{R}|\hat{u}_{R}=u_{R}+e_{R})p(e_{R})\right)\right\}^{2}.
\end{aligned}
\end{equation}
The average value of (\ref{GFq_SNR})
\begin{equation}\label{avgSNR_Qinv_GFq}
    \bar{\gamma}_{Q^{-1}}= \frac{1}{2\sin^{2}\left(\pi/M\right)} E\left[ \left\{ Q^{-1}(0.5 P_{Q^{-1}}^{s})\right\}^{2}\right]
\end{equation}
has no known closed-form expression but can easily be obtained numerically.

\begin{figure}[!t]
\centering
\includegraphics[width=4.5in]{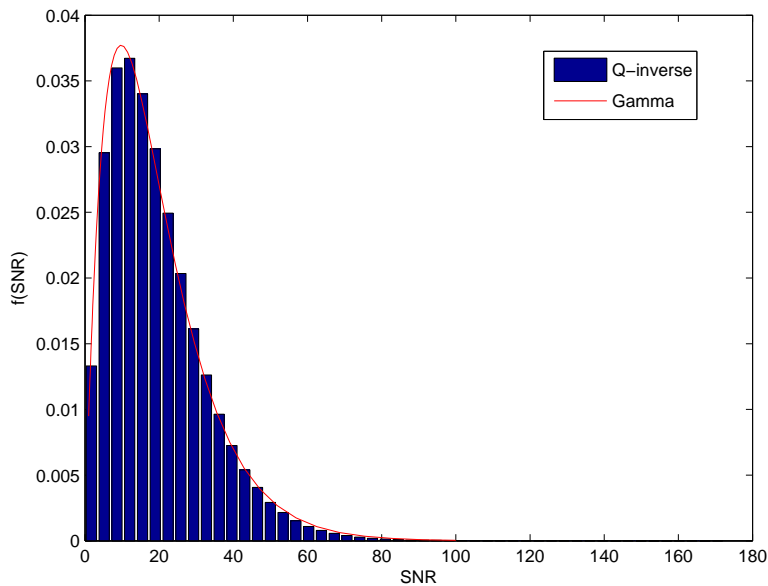}
\caption{Histogram of the Q-inverse equivalent channel's instantaneous SNR value in a Nakagami-m fading environment (m=2)}
\label{Qinv_hist}
\end{figure}

In a Nakagami-m fading environment, the equivalent SNR values given by (\ref{GFq_SNR}) are used in the Monte Carlo simulation of the Q-inverse equivalent receiver (see Fig. \ref{2user_GF2_Nakagami_hML}). The SNR values obtained via the Monte Carlo simulation  are shown in the histogram result in Fig. \ref{Qinv_hist}. It is seen that the distribution of $\gamma_{Q^{-1}}$ is well approximated by a Gamma distribution when $m=2$, so the equivalent relay channel can be approximated as a single-hop Nakagami-m channel. Based on this result, in a Nakagami-m fading environment, the instantaneous SNR value in (\ref{GFq_SNR}) is assumed to be Gamma distributed  \cite{Safak2013}:
\begin{equation}\label{fGFq_Qinv}
    f_{\gamma_{Q^{-1}}}(\gamma)=\frac{\gamma^{m-1}}{\Gamma(m)} \left(\frac{m}{\bar{\gamma}_{Q^{-1}}}\right)^{m} \exp\left(-\frac{m\gamma}{\bar{\gamma}_{Q^{-1}}}\right),\quad \gamma\geq0 .
\end{equation}

When soft-decision ML detection is performed at $D$, the likelihood function of the channel over which network coded data is sent is approximated with the likelihood function of a single-hop Q-inverse equivalent relay channel:
\begin{equation}\label{Q_Peq_sML}
 \begin{aligned}
  p_{Q^{-1}}(y_{D,nc}|u_{R})=\frac{1}{\pi \sigma^{2}_{Q^{-1}}}\exp\left(-\frac{1}{\sigma^{2}_{Q^{-1}}}\left|y_{D,nc}-h_{Q^{-1}}\varphi(u_R)\right|^2\right),
\end{aligned}
\end{equation}
where $h_{Q^{-1}}$ is the fading coefficient and $\sigma^{2}_{Q^{-1}}$ is the noise variance of the Q-inverse equivalent relay channel.

\emph{The Q-inverse equivalent soft-receiver} of the relay network in Fig. \ref{relay} obtains the ML estimates of $u_{i}$ with the soft-decision ML rule in (\ref{soft_MLD1}), where the likelihood functions of network coded data in  (\ref{py_u}) are approximated by (\ref{Q_Peq_sML}).

The Q-inverse equivalent soft-receiver is explained with the following example. Consider the BPSK modulated, 2-user GF(2) NC DetF relay network in Fig. \ref{relay} ($N=2$) which has a generator matrix given by (\ref{G2}). \begin{equation}\label{G2}
\mathbf{G_{2}}=\left(
                                         \begin{array}{cccc}
                                           1 & 0 & 1 & 1 \\
                                           0 & 1 & 1 & 1 \\
                                         \end{array}
                                       \right)
\end{equation}
Network coding is performed at $T_{1}$ and $T_{2}$ as $u_{1R}=u_{2R}=u_{1}+u_{2}$, respectively, where $u_{1}, u_{2}, u_{1R}, u_{2R} \in \mathrm{GF}(2)$. The received signals at $D$ are detected as in (\ref{yD}), where $N=2, K=4$. The likelihood functions of the detected data are used in the soft-decision ML rule in (\ref{soft_MLD1}), where the likelihood functions of the network coded data are approximated using the Q-inverse equivalent channel method as follows.
The likelihood function of the channel over which $u_{1R}$ is sent is approximated as
\begin{equation}\label{softReceiver_1}
 \begin{aligned}
  p_{Q^{-1},1}(y_{1D,nc}|u_{1R})=\frac{1}{\pi \sigma^{2}_{Q^{-1},1}}\exp\left(-\frac{1}{\sigma^{2}_{Q^{-1},1}}\left|y_{1D,nc}-h_{Q^{-1},1}\varphi(u_{1R})\right|^2\right).
\end{aligned}
\end{equation}
The fading coefficient $h_{Q^{-1},1}$ is obtained from the log-likelihood ratio (LLR) of $y_{1D,nc}$
\begin{equation}\label{LLR_y1D}
 \begin{aligned}
    L(y_{1D,nc}|u_{1R})&=\ln\left(\frac{p(y_{1D,nc}|u_{1R}=0)}{p(y_{1D,nc}|u_{1R}=1)}\right)\\
    &\simeq \frac{2\mathrm{Re}\{y_{1D,nc}h^*_{Q^{-1},1}\}}{\sigma^{2}_{Q^{-1},1}},
\end{aligned}
\end{equation}
where $\sigma^{2}_{Q^{-1},1}$ is equal to the variance of the noise term of $y_{1D,nc}$.
Similarly, the likelihood function of the channel over which $u_{2R}$ is sent is approximated as
\begin{equation}\label{softReceiver_2}
 \begin{aligned}
  p_{Q^{-1},2}(y_{2D,nc}|u_{2R})=\frac{1}{\pi \sigma^{2}_{Q^{-1},2}}\exp\left(-\frac{1}{\sigma^{2}_{Q^{-1},2}}\left|y_{2D,nc}-h_{Q^{-1},2}\varphi(u_{2R})\right|^2\right).
\end{aligned}
\end{equation}
The fading coefficient $h_{Q^{-1},2}$ is obtained using (\ref{LLR_y2D})
\begin{equation}\label{LLR_y2D}
 \begin{aligned}
    L(y_{2D,nc}|u_{2R})&=\ln\left(\frac{p(y_{2D,nc}|u_{2R}=0)}{p(y_{2D,nc}|u_{2R}=1)}\right)\\
    &\simeq \frac{2\mathrm{Re}\{y_{2D,nc}h^*_{Q^{-1},2}\}}{\sigma^{2}_{Q^{-1},2}},
\end{aligned}
\end{equation}
where $\sigma^{2}_{Q^{-1},2}$ is equal to the variance of the noise term of $y_{2D,nc}$.

\emph{The Q-inverse equivalent hard-receiver} of the relay network in Fig. \ref{relay} obtains the ML estimates of $u_{i}$ with the hard-decision ML rule in (\ref{hard_MLD}), where the SEP terms of network coded data in the likelihood function in (\ref{hMLD_CSI}) are approximated by (\ref{GFq_Peq}).

The Q-inverse equivalent hard-receiver is illustrated with an example as follows. Consider the QPSK modulated, 2-user GF(4) NC DetF relay network in Fig. \ref{relay} for $N=2$, which has a generator matrix:
\begin{equation}\label{G1}
\mathbf{G_{1}}=\left(
                                         \begin{array}{cccc}
                                           1 & 0 & 1 & 1 \\
                                           0 & 1 & 2 & 1 \\
                                         \end{array}
                                       \right).
\end{equation}
Hence, NC is performed at $T_{1}$ and $T_{2}$ as $u_{1R}=u_{1}+2u_{2}$ and $u_{2R}=u_{1}+u_{2}$, respectively, where $u_{1}, u_{2}, u_{1R}, u_{2R} \in \mathrm{GF}(4)$. The received data vector at $D$ is detected as in (\ref{Z}), where $N=2, K=4$.
The likelihood functions of the detected data are used in the hard-decision ML rule in (\ref{hard_MLD}), where the likelihood functions of the network coded data are approximated using the Q-inverse equivalent channel method as follows.
The instantaneous SEP of the relay channel $T_{2}-T_{1}-D$, over which $u_{1R}$ is sent, is approximated using (\ref{GFq_Peq}):
\begin{equation}\label{Peq_Qinv1}
\begin{aligned}
    P_{Q^{-1},1}^{s}&=2Q\left(\sqrt{\gamma_{Q^{-1},1}}\right)\\
    &= \sum_{e_{1R}=1}^{3} {p\left({z}_{1D,nc} \neq u_{1R}|\hat{u}_{1R}=u_{1R}+ e_{1R}\right) p( e_{1R})},
\end{aligned}
\end{equation}
where $e_{1R}$ is the network coded coherent QPSK demodulation error at $T_{1}$ and $\gamma_{Q^{-1},1}$ is the instantaneous SNR per symbol of the equivalent relay channel of $T_{2}-T_{1}-D$, which is found as
\begin{equation}\label{SNR_Qinv1}
    \gamma_{Q^{-1},1} = \{{Q^{-1}(0.5\, P_{Q^{-1},1}^{s})\}^{2}}.
\end{equation}
Similarly, the instantaneous SEP of the relay channel $T_{1}-T_{2}-D$, over which $u_{2R}$ is sent, is approximated
\begin{equation}\label{Peq_Qinv2}
\begin{aligned}
    P_{Q^{-1},2}^{s}&=2Q\left(\sqrt{\gamma_{Q^{-1},2}}\right)\\
    &=\sum_{e_{2R}=1}^{3} {p\left({z}_{2D,nc} \neq u_{2R}|\hat{u}_{2R}=u_{2R}+ e_{2R}\right) p(e_{2R})},
\end{aligned}
\end{equation}
where $e_{2R}$ is the network coded coherent QPSK demodulation error at $T_{2}$ and $\gamma_{Q^{-1},2}$ is the instantaneous SNR per symbol of the equivalent relay channel of $T_{1}-T_{2}-D$:
\begin{equation}\label{SNR_Qinv2}
    \gamma_{Q^{-1},2} = \{{Q^{-1}(0.5\, P_{Q^{-1},2}^{s})\}^{2}}.
\end{equation}

\subsubsection*{Minimum Equivalent Receiver}

Another equivalent relay channel model is the minimum equivalent relay channel model, which is commonly used in analyzing the error performance of DetF relay networks \cite{Gao2011,Yuan2010,Chu2008}. The minimum equivalent relay channel has a lower computational complexity than the Q-inverse equivalent relay channel, which is preferable in receiver design. In this model, the two-hop relay channel shown in Fig. \ref{equivalent} is approximated with a single-hop channel, where the equivalent instantaneous SNR  per symbol, $\gamma_{min}$, is chosen as the minimum of the $S-R$ and $R-D$ channels instantaneous SNR values \cite{Gao2011,Yuan2010,Chu2008}
\begin{equation}\label{snr_min}
    \gamma_{min}=\min\{\gamma_{SR},\gamma_{RD}\}.
\end{equation}
Here, $\gamma_{SR}$ and $\gamma_{RD}$ are the instantaneous SNR's per symbol of the $S-R$ and $R-D$ channels, respectively. In a Nakagami-m fading environment, the SNR value in (\ref{snr_min}) is distributed as \cite{Papoulis2002}
\begin{equation}\label{fmin_Nakagami}
\begin{aligned}
    f_{\gamma_{min}}(\gamma)&=\frac{1}{\Gamma(m_{SR})\, \gamma}\left(\frac{m_{SR} \, \gamma}{\bar{\gamma}_{SR}}\right)^{m_{SR}} e^ {-\left(\frac{m_{SR}}{\bar{\gamma}_{SR}}+\frac{m_{RD}}{\bar{\gamma}_{RD}}\right)\gamma} \sum_{k=0}^{m_{RD}-1}\frac{1}{k!}\left(\frac{m_{RD}\gamma}{\bar{\gamma}_{RD}}\right)^k \\
    & + \frac{1}{\Gamma(m_{RD})\, \gamma}\left(\frac{m_{RD} \, \gamma}{\bar{\gamma}_{RD}}\right)^{m_{RD}} e^ {-\left(\frac{m_{SR}}{\bar{\gamma}_{SR}}+\frac{m_{RD}}{\bar{\gamma}_{RD}}\right)\gamma}     \sum_{k=0}^{m_{SR}-1}\frac{1}{k!}\left(\frac{m_{SR}\gamma}{\bar{\gamma}_{SR}}\right)^k,
\end{aligned}
\end{equation}
where $m_{SR}$ and $m_{RD}$ are the fading figures of the $S-R$ and $R-D$ links  respectively; $\bar{\gamma}_{SR}$ and $\bar{\gamma}_{RD}$ are the average SNR values of the $S-R$ and $R-D$ links, respectively. When $m_{SR}=m_{RD}=m$, the distribution in (\ref{fmin_Nakagami}) is reduced to
\begin{equation}\label{fmin_Nakagami_m}
\begin{aligned}
    f_{\gamma_{min}}(\gamma)&=\frac{\left(m \gamma\right)^{m}}{\Gamma(m) \gamma} e^ {-\left(\frac{1}{\bar{\gamma}_{SR}}+\frac{1}{\bar{\gamma}_{RD}}\right)m \gamma} \\ & \times \left\{ \frac{1}{(\bar{\gamma}_{SR})^{m}}\sum_{k=0}^{m-1}\frac{1}{k!}\left(\frac{m\gamma}{\bar{\gamma}_{RD}}\right)^k+    \frac{1}{(\bar{\gamma}_{RD})^{m}}\sum_{k=0}^{m-1}\frac{1}{k!}\left(\frac{m\gamma}{\bar{\gamma}_{SR}}\right)^k
    \right\}.
\end{aligned}
\end{equation}
The average value of (\ref{snr_min}) when $m_{SR}=m_{RD}=m$ is found as
\begin{equation}\label{Emin}
\begin{aligned}
    \bar\gamma_{min}&=-2 (-1)^{-m} \frac{ \Gamma(2m)}{\Gamma(m)^2}
    \left\{ \bar{\gamma}_{RD} \,{\rm B}\left(-\frac{\bar{\gamma}_{SR}}{\bar{\gamma}_{RD}}, m+1, -2m\right)\right. \\
    &\left.\quad +\,\bar{\gamma}_{SR}\, {\rm B}\left(-\frac{\bar{\gamma}_{RD}}{\bar{\gamma}_{SR}}, m+1, -2m\right)\right\},
\end{aligned}
\end{equation}
where ${\rm B}(x,y,z)$ is the incomplete Beta function.

The minimum equivalent relay of the N-source relay channel in Fig. \ref{GFq_equivalent} has an instantaneous SNR value
\begin{equation}\label{SNRmin_Nuser}
    \gamma_{min}=\min\{\gamma_{S_{1}R},\gamma_{S_{2}R}, \ldots \gamma_{S_{N}R},\gamma_{RD}\},
\end{equation}
where $\gamma_{S_{n}R}$ is the instantaneous SNR per symbol of the channel $S_{n}-R$, $1 \leq n \leq N$.
The distribution of (\ref{SNRmin_Nuser}) and its average SNR value, however, is not as readily found for large values of $N$, making the model unsuitable for error rate analysis but useful for designing a low-complexity receiver, which is described below.

When soft-decision ML detection is performed at $D$, the likelihood function of the channel over which network coded data is sent is approximated with the likelihood function of a single-hop Q-inverse equivalent relay channel:
\begin{equation}\label{min_Peq_sML}
 \begin{aligned}
  p_{min}(y_{D,nc}|u_{R})=\frac{1}{\pi \sigma^{2}_{min}}\exp\left(-\frac{1}{\sigma^{2}_{min}}\left|y_{D,nc}-h_{min}\varphi(u_R)\right|^2\right),
\end{aligned}
\end{equation}
where $h_{min}$ is the fading coefficient and $\sigma^{2}_{min}$ is the noise variance of the minimum equivalent relay channel and are found using (\ref{SNRmin_Nuser}), where $\gamma_{min}=\frac{E_S}{\sigma^{2}_{min}}|h_{min}|^2$.

\emph{The minimum equivalent soft-receiver} of the relay network in Fig. \ref{relay} obtains the ML estimates of $u_{i}$ with the soft-decision ML rule in (\ref{soft_MLD1}), where the likelihood functions of network coded data in  (\ref{py_u}) are approximated by (\ref{min_Peq_sML}). \emph{The minimum equivalent hard-receiver} of the relay network in Fig. \ref{relay} obtains the ML estimates of $u_{i}$ by the hard-decision ML rule in (\ref{hard_MLD}), where the SEP terms of the network coded data in the likelihood function in (\ref{hMLD_CSI}) are approximated to the coherent $M$-PSK demodulation SEP of a minimum equivalent relay channel which has an instantaneous SNR value per symbol $\gamma_{min}$:
\begin{equation}\label{hMLD_min}
\begin{aligned}
    &\sum_{e_{R}=1}^{q-1} {p\left(z_{D,nc}|\hat{u}_{R}=u_{R}+ e_{R}\right) p(e_{R})}\\
&    \simeq P_{min}^{s} = Q(\sqrt{2 \gamma_{min}}) + \frac{2}{\pi} \int_{0}^{\infty} \exp{\left[-\left( u-\sqrt{\gamma_{min}}\right)^{2}\right]}Q\left(\sqrt{2}u \tan(\pi/M)\right)\mathrm{d}u.
\end{aligned}
\end{equation}

For analytical tractability in the error rate analysis, the minimum and Q-inverse equivalent relay channel models will be used in the error rate analysis, which will be discussed in Section \ref{Error Rate Analysis}.

\section{Error Rate Analysis}\label{Error Rate Analysis}

\subsection{Relay Networks with Error-Free Source-Relay Channels}
The error rate analysis of the optimum soft-receiver of the relay network with error-free $S-R$ links, which is described in Section \ref{section:optimum}, is provided as a preliminary to the error rate analysis of the relay network with error-prone $S-R$ links.

In the error rate analysis of the relay network with error-free $S-R$ links, the reliability of the observation at the destination depends only on the $R-D$ channel SNR values. The SNR vector observed at $D$ consists of the instantaneous SNR values of (\ref{y}):
\begin{equation}\label{SNR_vector}
\begin{array}{cccccc}
\mathbf{\Gamma}=[  \gamma_{1D,sys} & \ldots & \gamma_{ND,sys} & \gamma_{1D,nc} & \ldots & \gamma_{(K-N)D,nc}],
\end{array}
\end{equation}
where $\gamma_{nD,sys}, 1\leq n \leq N,$  is the instantaneous SNR value per symbol of the $S_n-R$ channel over which $u_{n}$ is transmitted and $\gamma_{mD,nc}, 1\leq m \leq K-N,$ is the instantaneous SNR value per symbol of the $R-D$ channel over which $u_{mR}$ is sent. The average SNR vector observed at $D$ comprises the expected values of the instantaneous SNR values in (\ref{SNR_vector})
\begin{equation}\label{avgSNR_vector}
\begin{array}{cccccc}
\mathbf{\bar{\Gamma}}=[  \bar{\gamma}_{1D,sys} & \ldots & \bar{\gamma}_{ND,sys} & \bar{\gamma}_{1D,nc} & \ldots & \bar{\gamma}_{(K-N)D,nc}].
\end{array}
\end{equation}
The transmitted block in Euclidean space is the modulated codeword in (\ref{X}). The SEP is upper bounded by the average  of the average pairwise error probabilities (PEPs) \cite{Simon2000}
\begin{equation}\label{union}
P_{S}\leq\frac{1}{K\left\vert \chi \right\vert}
\displaystyle\sum\limits_{k=1}^{K}
\displaystyle\sum\limits_{ \mathbf{X}(i), \mathbf{\hat{X}}(j)  \epsilon
\chi ,j_{k}\neq i_{k}}
P\left(  \mathbf{X}(i)  \rightarrow\mathbf{\hat{X}}(j)  \right),
\end{equation}
where $\mathbf{\hat{X}}(j)$ is the decision at $D$; $\mathbf{X}(i)$ and $\mathbf{\hat{X}}(j)$ are found using (\ref{X}).
Note that each PEP, $P\left(  \mathbf{X}(i)  \rightarrow\mathbf{\hat{X}}(j)  \right)$, in (\ref{union}) is obtained after averaging over the channel matrix $\mathbf{H_{D}}$. This averaging can be performed using the moment generating function (MGF) method \cite{Simon2000,Goldsmith2005}:
\begin{equation}\label{MGF}
P\left(\mathbf{X}(i)  \mathbf{\rightarrow\mathbf{\hat{X}}}(j)\right) =\frac{1}{\pi}
\int\limits_{\theta=0}^{\pi/2}\prod_{k=1}^{K}M_{k}\left(-\frac{\left|\varphi(i_{k})-\varphi(j_{k})\right|^{2}}{4\sin^{2}(\theta)}\right)\mathrm{d}\theta.
\end{equation}
Here, $M_{k}(s)$ is the MGF of the $k^{th}$ element of (\ref{SNR_vector}). For Nakagami-m fading, the MGF is $M_{k}(s)=(1-s\bar{\gamma}_{k}/m)^{-m}$,  where $m$ is the fading figure and $\bar{\gamma}_{k}$ is the $k^{th}$ element of  (\ref{avgSNR_vector}) \cite{Papoulis2002,Proakis2000}. In a Nakagami-m fading environment, the PEP in (\ref{MGF}) is expanded as
\begin{equation}\label{MGF_1}
\begin{aligned}
P\left(\mathbf{X}(i)  \mathbf{\rightarrow\mathbf{\hat{X}}}(j)\right)
&=\frac{1}{\pi}
\int\limits_{\theta=0}^{\pi/2}\prod_{k=1}^{K}\left(\frac{4m\sin^{2}(\theta)}{4m\sin^{2}(\theta)+\bar{\gamma}_{k}\left|\varphi(i_{k})-\varphi(j_{k})\right|^{2}}\right)^{m}\mathrm{d}\theta\\
&=\frac{1}{\pi}
\int\limits_{\theta=0}^{\pi/2}
\frac{\left(4m\sin^{2}(\theta)\right)^{mK}}{\displaystyle\prod_{k=1}^{N}\left(4m\sin^{2}(\theta)+\bar{\gamma}_{kD,sys}\left|\varphi(i_{k})-\varphi(j_{k})\right|^{2}\right)^m}
\\&\times\frac{1}{\displaystyle\prod_{l=1}^{K-N}\left(4m\sin^{2}(\theta)+\bar{\gamma}_{lD,nc}\left|\varphi(i_{N+l})-\varphi(j_{N+l})\right|^{2}\right)^m} \, \mathrm{d}\theta.
\end{aligned}
\end{equation}
The diversity order, $DO$, is found with the help of (\ref{union}) \cite{Liu2008}:
\begin{equation}\label{diversity}
DO \triangleq \lim_{\gamma \rightarrow \infty} \frac{1}{\log(\gamma)\, K\left\vert \chi \right\vert}
\displaystyle\sum\limits_{k=1}^{K}
\displaystyle\sum\limits_{ \mathbf{X}(i), \mathbf{\hat{X}}(j)  \epsilon
\chi ,j_{k}\neq i_{k}}
P\left(  \mathbf{X}(i)  \rightarrow\mathbf{\hat{X}}(j)  \right),
\end{equation}
where $\gamma$ is the average SNR value of the relay network assuming the average SNR values in (\ref{avgSNR_vector}) are all equal.
Assuming $\bar{\gamma}_{kD,sys}\left|\varphi(i_{k})-\varphi(j_{k})\right|^{2}\gg 4m\sin^{2}(\theta) $ and $\bar{\gamma}_{lD,nc}\left|\varphi(i_{N+l})-\varphi(j_{N+l})\right|^{2} \gg 4m\sin^{2}(\theta)$ for high SNR values, the asymptotic values of the PEPs in (\ref{diversity}) are obtained:
\begin{equation}\label{MGF_highSNR}
\begin{aligned}
&P\left(\mathbf{X}(i)  \mathbf{\rightarrow\mathbf{\hat{X}}}(j)\right)
\leq \frac{1}{\pi}
\int\limits_{\theta=0}^{\pi/2}
\frac{\left(4m\sin^{2}(\theta)\right)^{m(K-M)}}{\displaystyle\prod_{k=1,i_{k} \neq j_{k}}^{N}(\bar{\gamma}_{kD,sys})^m\left|\varphi(i_{k})-\varphi(j_{k})\right|^{2m}}
\\&\times\frac{1}{\displaystyle\prod_{l=1, i_{l} \neq j_{l}}^{K-N}(\bar{\gamma}_{lD,nc})^m\left|\varphi(i_{N+l})-\varphi(j_{N+l})\right|^{2m}} \, \mathrm{d}\theta\\
&=\frac{(4m)^{m(K-M)} \displaystyle \int\limits_{\theta=0}^{\pi/2}
\sin^{2m(K-M)}(\theta) \mathrm{d}\theta}{\pi\displaystyle\prod_{k=1,i_{k} \neq j_{k}}^{N}(\bar{\gamma}_{kD,sys})^m\left|\varphi(i_{k})-\varphi(j_{k})\right|^{2m}\displaystyle\prod_{l=1, i_{l} \neq j_{l}}^{K-N}(\bar{\gamma}_{lD,nc})^m\left|\varphi(i_{N+l})-\varphi(j_{N+l})\right|^{2m} },
\end{aligned}
\end{equation}
where $M$ is the total number of elements $k$ which satisfy $i_{k} \neq j_{k}, 1 \leq k \leq K$. Using Formula 3.621.3 on p.389 in \cite{Gradshteyn2000}, the integral term in (\ref{MGF_highSNR}) is found as
\begin{equation}\label{sin_integral}
\int\limits_{\theta=0}^{\pi/2}
\sin^{2m(K-M)}(\theta) \mathrm{d}\theta= \frac{\pi}{2}\frac{(2m[K-M]-1)!!}{(2m[K-M])!!},
\end{equation}
where $!!$ denotes double factorial \cite{Gradshteyn2000}. Inserting (\ref{sin_integral}) into (\ref{MGF_highSNR}) gives
\begin{equation}\label{PEP_highSNR}
\begin{aligned}
&P\left(\mathbf{X}(i)  \mathbf{\rightarrow\mathbf{\hat{X}}}(j)\right)
\leq
 \displaystyle \frac{(4m)^{m(K-M)}(2m[K-M]-1)!!}{2(2m[K-M])!!}\\&\times
\displaystyle\left(\prod_{k=1,i_{k} \neq j_{k}}^{N}(\bar{\gamma}_{kD,sys})^m\left|\varphi(i_{k})-\varphi(j_{k})\right|^{2m}\displaystyle\prod_{k=1, i_{k} \neq j_{k}}^{K-N}(\bar{\gamma}_{kD,nc})^m\left|\varphi(i_{N+k})-\varphi(j_{N+k})\right|^{2m}\right)^{-1}.
\end{aligned}
\end{equation}

\subsection{Relay Networks with Error-Prone Source-Relay Channels}

For the relay networks with error-prone $S-R$ channels, the error rate analysis is not as simple as in the error-free case. This is due to two facts. First, the optimal soft-receiver is complex because of the averaging of $S-R$ link errors, as shown in (\ref{py_u}). Second, even if a suboptimal receiver is used which does not perform this averaging, the analysis is still complex due to the $S-R$ error events. In order to obtain a tractable analytical results for the PEPs, we propose using equivalent relay channel models, which are described in Section \ref{section:equivalent}, for the analysis of the optimum soft-receiver. Thus, we will approximate the error-prone relay channel with a single-hop equivalent relay channel in order to obtain the approximate error rate and diversity order expressions of the optimum soft-receiver.

An approximate error rate of the relay network in Fig. \ref{relay} is found using an equivalent relay channel model in the following manner. The channel over which the network coded data at the terminals are sent to the destination are approximated using an equivalent channel model. The average PEPs of the union bound in (\ref{union}) are approximated using the same method employed in the error-free case, given by (\ref{MGF}), except that the SNR vector is obtained using the equivalent relay channels:
\begin{equation}\label{SNR_vector_eq}
\begin{array}{cccccc}
\mathbf{\Gamma}_{eq}=[  \gamma_{1D,sys} & \ldots & \gamma_{ND,sys} & \gamma_{eq,1} & \ldots & \gamma_{eq,K-N}]
\end{array}
\end{equation}
where $\gamma_{nD,sys}, 1\leq n \leq N,$  is the instantaneous SNR value per symbol of the channel over which $u_{n}$ is transmitted and $\gamma_{eq,j}, 1\leq j \leq K-N,$ is the instantaneous SNR value per symbol of the equivalent relay channel over which $u_{jR}$ is sent. The average SNR vector is the found by taking the expected value of each instantaneous SNR value in (\ref{SNR_vector_eq})
\begin{equation}\label{avgSNR_vector_eq}
\begin{array}{cccccc}
\mathbf{\bar{\Gamma}}_{eq}=[  \bar{\gamma}_{1D,sys} & \ldots & \bar{\gamma}_{ND,sys} & \bar{\gamma}_{eq,1} & \ldots & \bar{\gamma}_{eq,K-N}].
\end{array}
\end{equation}
Hence, the relay network is approximated with an equivalent relay network.
The PEPs of the equivalent relay network are found by letting (\ref{avgSNR_vector_eq}) into (\ref{MGF}):
\begin{equation}\label{MGF_eq}
\begin{aligned}
P\left(\mathbf{X}(i)  \mathbf{\rightarrow\mathbf{\hat{X}}}(j)\right)
&=\frac{1}{\pi}
\int\limits_{\theta=0}^{\pi/2}
\frac{\left(4m\sin^{2}(\theta)\right)^{mK}}{\displaystyle\prod_{k=1}^{N}\left(4m\sin^{2}(\theta)+\bar{\gamma}_{kD,sys}\left|\varphi(i_{k})-\varphi(j_{k})\right|^{2}\right)^m}
\\&\times \frac{1}{\displaystyle\prod_{k=1}^{K-N}\left(4m\sin^{2}(\theta)+\bar{\gamma}_{eq,k}\left|\varphi(i_{N+k})-\varphi(j_{N+k})\right|^{2}\right)^m} \, \mathrm{d}\theta.
\end{aligned}
\end{equation}
The SEP of the equivalent relay network is then obtained by inserting (\ref{MGF_eq}) into (\ref{union}). The diversity order of the equivalent relay network is found using (\ref{diversity}), where
the asymptotic PEPs of the equivalent relay network are used:
\begin{equation}\label{MGFeq_highSNR}
\begin{aligned}
&P\left(\mathbf{X}(i)  \mathbf{\rightarrow\mathbf{\hat{X}}}(j)\right)
\leq \frac{1}{\pi}
\int\limits_{\theta=0}^{\pi/2}
\frac{\left(4m\sin^{2}(\theta)\right)^{m(K-M)}}{\displaystyle\prod_{k=1,i_{k} \neq j_{k}}^{N}(\bar{\gamma}_{kD,sys})^m\left|\varphi(i_{k})-\varphi(j_{k})\right|^{2m}}
\\&\times\frac{1}{\displaystyle\prod_{k=1, i_{k} \neq j_{k}}^{K-N}(\bar{\gamma}_{eq,k})^m\left|\varphi(i_{N+k})-\varphi(j_{N+k})\right|^{2m}} \, \mathrm{d}\theta\\
&=\frac{(4m)^{m(K-M)} \displaystyle \frac{(2m[K-M]-1)!!}{2(2m[K-M])!!}}{\displaystyle\prod_{k=1,i_{k} \neq j_{k}}^{N}(\bar{\gamma}_{kD,sys})^m\left|\varphi(i_{k})-\varphi(j_{k})\right|^{2m}\displaystyle\prod_{k=1, i_{k} \neq j_{k}}^{K-N}(\bar{\gamma}_{eq,k})^m\left|\varphi(i_{N+k})-\varphi(j_{N+k})\right|^{2m} }.
\end{aligned}
\end{equation}

Let us illustrate how the PEPs of the Q-inverse and minimum equivalent relay networks are obtained with an example. Consider the 2-user GF(4) NC DetF, QPSK modulated relay network  described in (\ref{G1})-(\ref{SNR_Qinv2}) assuming a Nakagami-m fading environment with $m=1$ (Rayleigh flat-fading). The relay network has a generator matrix given by (\ref{G1}) and the equivalent relay network's SNR vector at $D$ is given by (\ref{SNR_vector_eq}), where $N=2,\,K=4$. Assuming the $S-R$ and $R-D$ channels have equal average SNR values, the average SNR values in (\ref{avgSNR_vector_eq}) are $\bar{\gamma}_{1D,sys} = \bar{\gamma}_{2D,sys}= \bar{\gamma}$ and $\bar{\gamma}_{eq,1} = \bar{\gamma}_{eq,2}=\bar{\gamma}_{eq}$ and the equivalent relay network's average SNR vector is
 \begin{equation}\label{avgSNR}
\begin{array}{cccccc}
\mathbf{\bar{\Gamma}}_{eq}=[ \bar{\gamma} & \bar{\gamma} & \bar{\gamma}_{eq} & \bar{\gamma}_{eq}].
\end{array}
\end{equation}
Inserting $m=1$ and $\bar{\gamma}_{SR}=\bar{\gamma}_{RD}=\bar{\gamma}$ in (\ref{Emin}), the minimum equivalent relay channel's average SNR value is found as $\bar{\gamma}_{min}=\bar{\gamma}/2$, hence the average SNR vector of the minimum equivalent relay network is the following:
\begin{equation}\label{avgSNR_min}
\begin{array}{cccccc}
\mathbf{\bar{\Gamma}}_{min}=[ \bar{\gamma} & \bar{\gamma} & \bar{\gamma}/2 & \bar{\gamma}/2].
\end{array}
\end{equation}
The average SNR vector of the Q-inverse equivalent relay network is given below:
\begin{equation}\label{avgSNR_Qinv}
\begin{array}{cccccc}
\mathbf{\bar{\Gamma}}_{Q^{-1}}=[ \bar{\gamma} & \bar{\gamma} & \bar{\gamma}_{Q^{-1}} & \bar{\gamma}_{Q^{-1}}]
\end{array}
\end{equation}
where $\bar{\gamma}_{Q^{-1}}$ is found by inserting (\ref{SNR_Qinv1}) and (\ref{SNR_Qinv2}) into (\ref{avgSNR_Qinv_GFq}) for $M=4$. As an example, the average PEP between the codewords $\begin{array}{cccc}
                                      \mathbf{X}(0)=\varphi([0 & 0 & 0 & 0])
                                    \end{array}$ and $\begin{array}{cccc}
                                      \hat{\mathbf{X}}(1)=\varphi([0 & 1 & 2 & 1])
                                    \end{array}$ is found using (\ref{MGF_eq}):
\begin{equation} \label{MGF_PEP}
\begin{aligned}
    P\left(\mathbf{X}(0)  \mathbf{\rightarrow\mathbf{\hat{X}}}(1)\right) &=\frac{1}{\pi}
    \int\limits_{\theta=0}^{\pi/2}\left(1+\bar{\gamma}\frac{\left|\varphi(0)-\varphi(0)\right|^{2}}{4\sin^{2}(\theta)}\right)^{-1}
    \left(1+\bar{\gamma}\frac{\left|\varphi(0)-\varphi(1)\right|^{2}}{4\sin^{2}(\theta)}\right)^{-1}\\
    &\quad\qquad\times \left(1+\bar{\gamma}_{eq}\frac{\left|\varphi(0)-\varphi(2)\right|^{2}}{4\sin^{2}(\theta)}\right)^{-2}\mathrm{d}\theta\\
    &=\frac{1}{\pi}\int\limits_{\theta=0}^{\pi/2}\left(\frac{2\sin^{2}(\theta)}{2\sin^{2}(\theta)+\bar{\gamma}}\right) \left(\frac{2\sin^{2}(\theta)}{2\sin^{2}(\theta)+\bar{\gamma}_{eq}}\right)^{2}\mathrm{d}\theta
\end{aligned}
\end{equation}
where the modulated codewords are taken as $\varphi(0)=1$,  $\varphi(1)=j$, $\varphi(2)=-j$ and $\varphi(3)=-1$, assuming Gray coding.
When the minimum equivalent relay network's SNR vector in (\ref{avgSNR_min}) is used in (\ref{MGF_PEP}), the PEP reduces to
\begin{equation} \label{MGF_min}
\begin{aligned}
    P\left(\mathbf{X}(0)  \mathbf{\rightarrow\mathbf{\hat{X}}}(1)\right) &=\frac{1}{\pi}\int\limits_{\theta=0}^{\pi/2}\left(\frac{2\sin^{2}(\theta)}{2\sin^{2}(\theta)+\bar{\gamma}}\right) \left(\frac{4\sin^{2}(\theta)}{4\sin^{2}(\theta)+\bar{\gamma}}\right)^{2}\mathrm{d}\theta.
\end{aligned}
\end{equation}
Inserting the Q-inverse equivalent relay network's SNR vector in (\ref{avgSNR_Qinv}) into (\ref{MGF_PEP}) gives
\begin{equation} \label{MGF_Qinv}
\begin{aligned}
    P\left(\mathbf{X}(0)  \mathbf{\rightarrow\mathbf{\hat{X}}}(1)\right) &=\frac{1}{\pi}\int\limits_{\theta=0}^{\pi/2}\left(\frac{2\sin^{2}(\theta)}{2\sin^{2}(\theta)+\bar{\gamma}}\right) \left(\frac{2\sin^{2}(\theta)}{2\sin^{2}(\theta)+\bar{\gamma}_{Q^{-1}}}\right)^{2}\mathrm{d}\theta.
 \end{aligned}
\end{equation}
For high SNR values, using the fact that the Q-inverse equivalent relay channel converges to the minimum equivalent relay channel \cite{Wang2007,Nasri2010}, the union bound from (\ref{MGF_min}) and (\ref{MGF_Qinv}) is found using (\ref{sin_integral}):
\begin{equation} \label{PEP01}
\begin{aligned}
    P\left(\mathbf{X}(0)  \mathbf{\rightarrow\mathbf{\hat{X}}}(1)\right)
    & \leq  \frac{32}{\pi \bar{\gamma}^{3}}
    \int\limits_{\theta=0}^{\pi/2}{\sin^{6}(\theta)}\mathrm{d}\theta
    = \frac{32}{\pi\bar{\gamma}^{3}} \frac{\pi\, 5!!}{2 \,6!!}
    = \frac{5}{\bar{\gamma}^{3}}.
\end{aligned}
\end{equation}
Similarly, the other PEP values are also found for high SNR values, which are then used in obtaining the diversity order given in (\ref{diversity}). The dominating term in this case is the PEP value in (\ref{PEP01}), hence the diversity order is equal to 3.

\section{Numerical Results}

In this section, the optimal receiver's error performance of the relay network shown in Fig. \ref{relay} is compared to the error performances of the minimum and Q-inverse equivalent receivers and the approximate union bounds obtained using the equivalent relay channel models. For each SNR value, at least 50 bit/symbol errors were observed in obtaining the corresponding BER/SER.
\begin{figure}[h]
\centering
\includegraphics[width=4.5in]{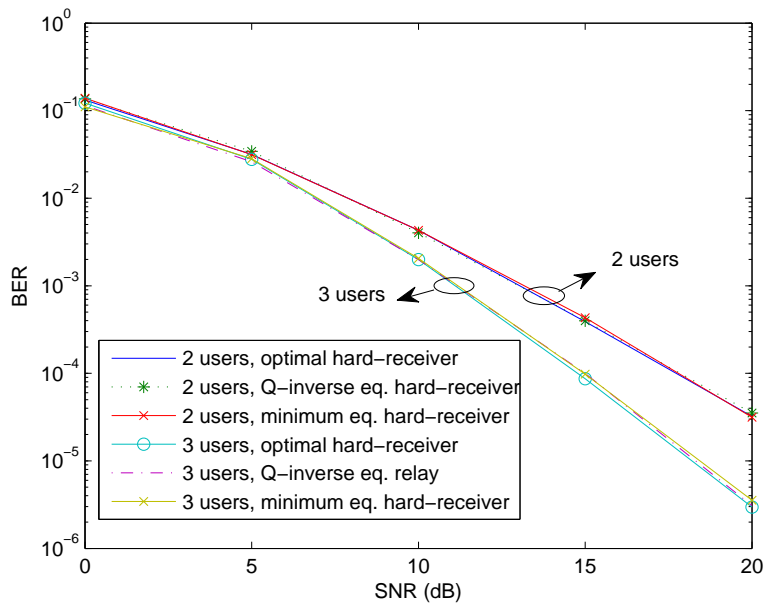}
\caption{2 and 3-user GF(2) NC DetF Nakagami-m (m=1) fading relay networks' hard-receiver performances}
\label{2_3user_GF2_hML}
\end{figure}

Fig. \ref{2_3user_GF2_hML} compares the optimum hard-receivers' BER performances of the 2- and 3-user cases of the GF(2) NC DetF Nakagami-m fading with $m=1$  relay network in Fig. \ref{relay} with that of the minimum and Q-inverse equivalent hard-receivers. The generator matrices of the 2- and 3-user relay networks are given in (\ref{G2}) and (\ref{G3}), respectively.

\begin{equation}\label{G3}
   \mathbf{G_{3}}= \left(
  \begin{array}{cccccc}
    1 & 0 & 0 & 1 & 1 & 1 \\
    0 & 1 & 0 & 1 & 1 & 0 \\
    0 & 0 & 1 & 1 & 0 & 1 \\
  \end{array}
\right)
\end{equation}
\begin{figure}[h!]
\centering
\includegraphics[width=4.5in]{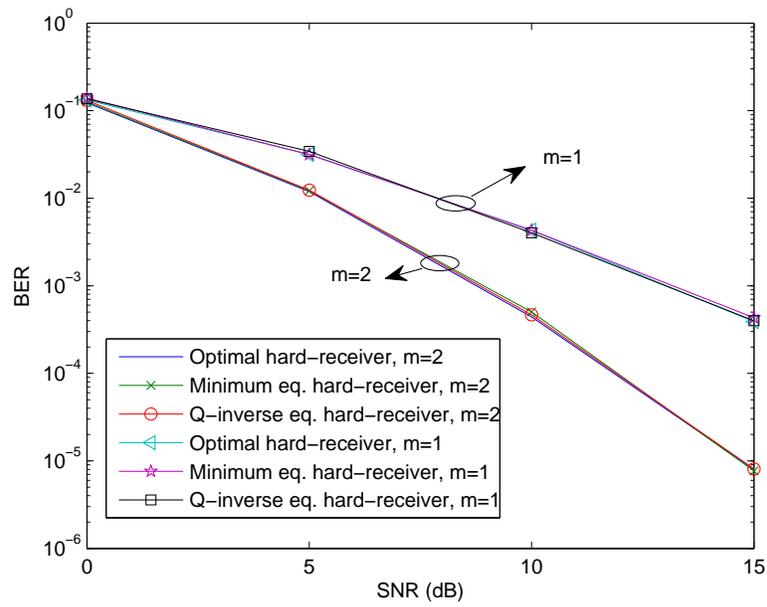}
\caption{2-user GF(2) NC DetF Nakagami-m fading relay network for m=1 and m=2,  hard-receiver performances}
\label{2user_GF2_Nakagami_hML}
\end{figure}
\begin{figure}[h!]
\centering
\includegraphics[width=4.5in]{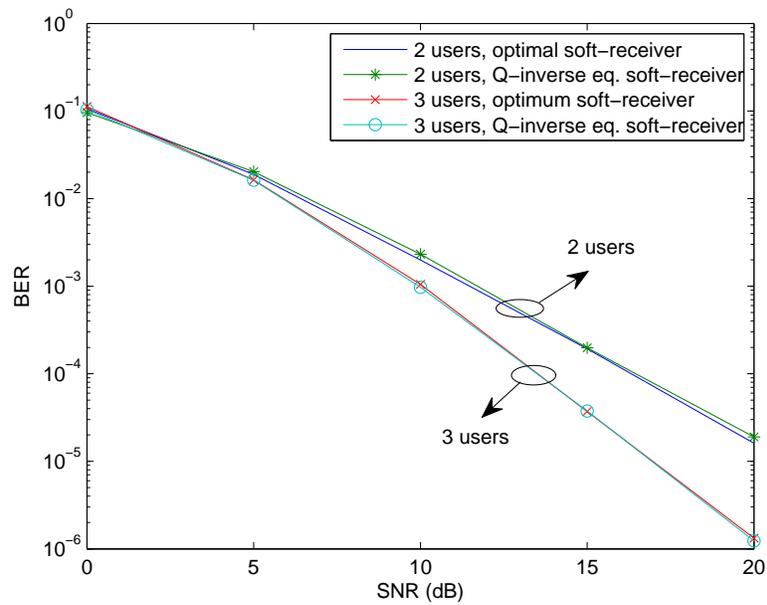}
\caption{2- vs 3-user GF(2) NC DetF Nakagami-m (m=1) fading relay networks' soft receiver performances}
\label{2_3user_GF2_softML}
\end{figure}
\begin{figure}[h!]
\centering
\includegraphics[width=4.5in]{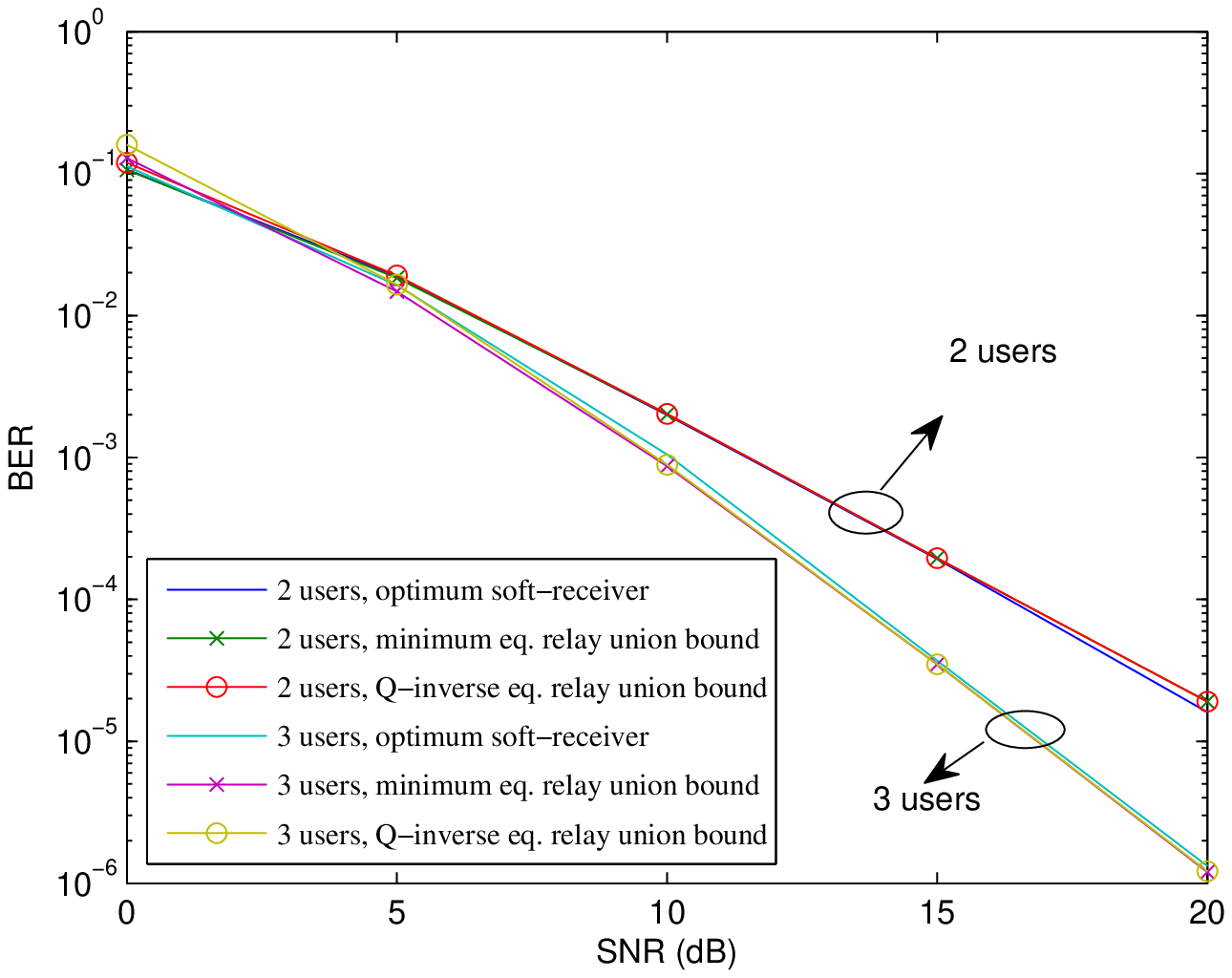}
\caption{2- and 3-user GF(2) NC DetF Nakagami-m (m=1) fading relay networks' optimum soft-receiver BER performances vs minimum and Q-inverse equivalent relay networks' union bound BER performances}
\label{2_3user_GF2_union}
\end{figure}
The Q-inverse and minimum equivalent hard-receivers are observed to display an error performance very close to each other and to the optimal hard-receiver. The diversity order is shown to increase with an increasing number of users; the diversity orders of the 2 and 3-user relay networks are 2 and 3, respectively. Hence, the equivalent receivers exhibit very close error performance to the optimal receiver independent of the number of users in the relay network.
\begin{figure}[h!]
\centering
\includegraphics[width=4.5in]{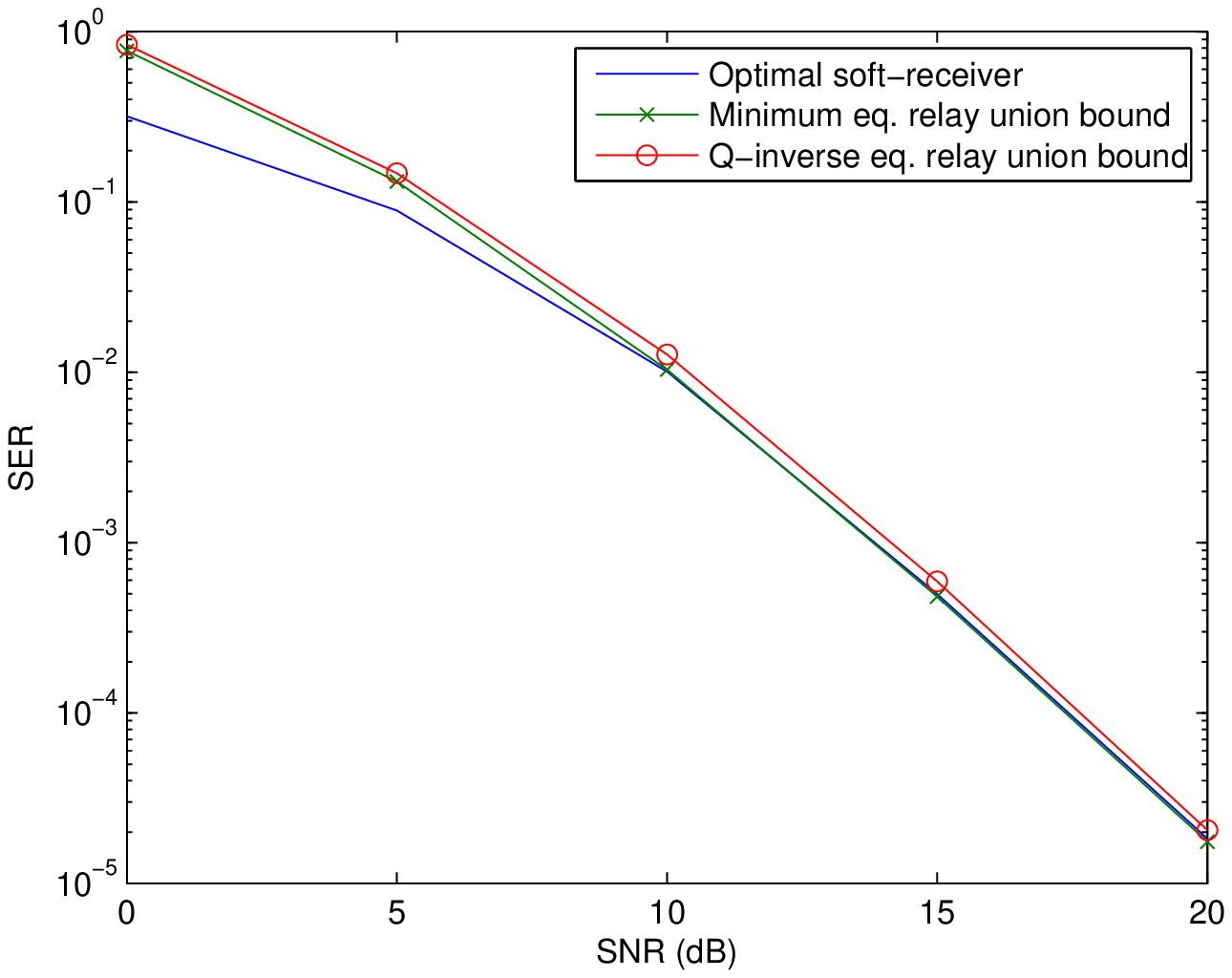}
\caption{2-user GF(4) NC DetF Nakagami-m (m=1) fading relay network's optimal soft-receiver SER vs minimum and Q-inverse equivalent relay networks' union bound SER performances}
\label{2user_GF4_union}
\end{figure}
\begin{figure}[h!]
\centering
\includegraphics[width=4.5in]{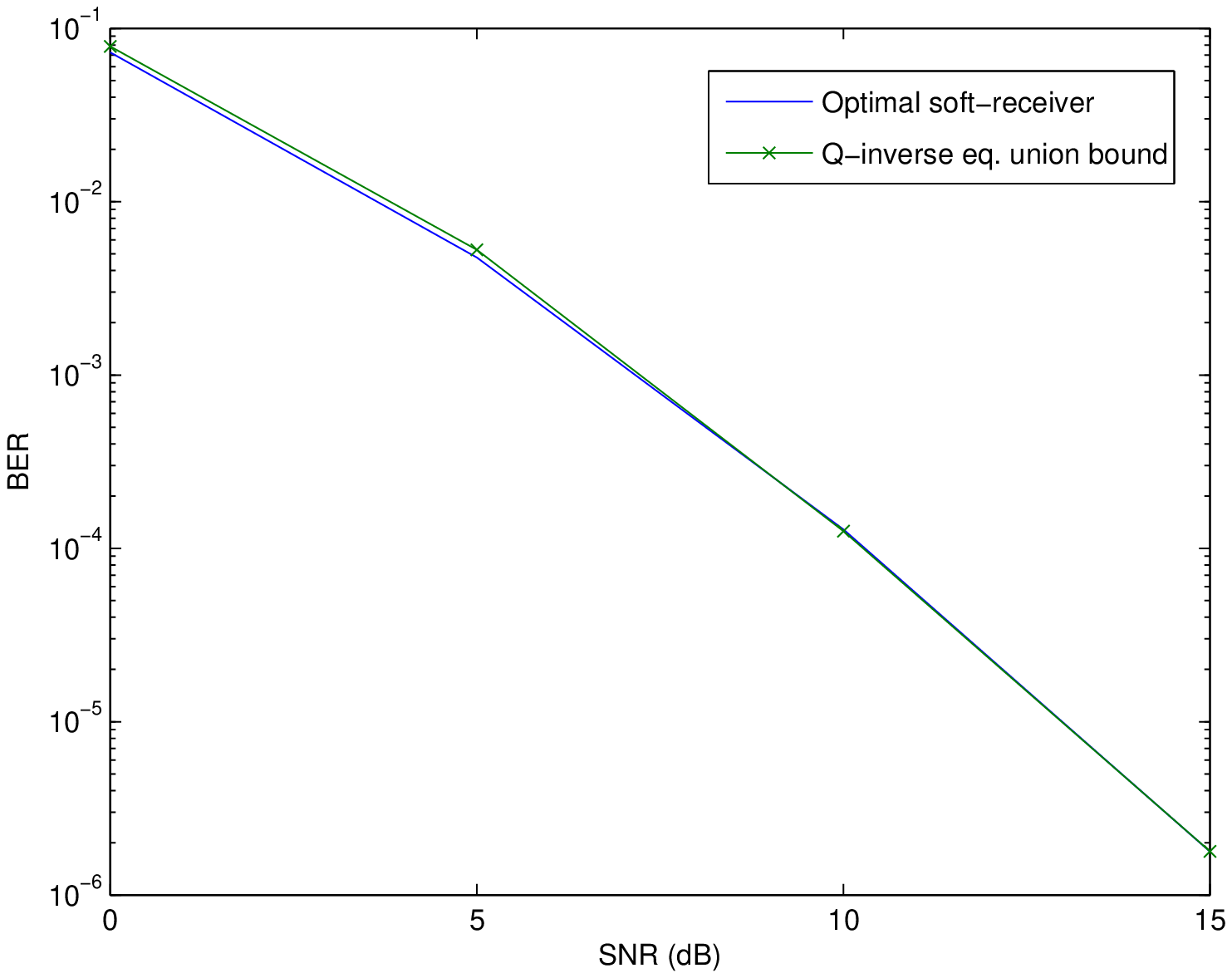}
\caption{2-user GF(2) NC DetF Nakagami-m (m=2) fading relay network's optimal soft-receiver BER vs Q-inverse equivalent relay network's union bound BER performance}
\label{GF2_Nakagami_union}
\end{figure}

Fig. \ref{2user_GF2_Nakagami_hML} compares the BER performance of the optimal hard-receiver of the 2-user GF(2) NC DetF Nakagami-m fading relay network in Fig. \ref{relay} with that of the minimum and Q-inverse equivalent hard-receivers for $m=1$ and $m=2$. The generator matrix of the relay network is given in (\ref{G2}). It is observed that the minimum and Q-inverse equivalent hard-receivers display very close error performances to the optimal receiver and have a diversity order close to 3 when $m=2$, compared to a diversity order of 2 when $m=1$. Hence, the equivalent receivers display very close error performances, independent of the fading figure.

Fig. \ref{2_3user_GF2_softML} compares the optimum soft-receiver BER performance of the of the 2- and 3-user GF(2) NC DetF Nakagami-m fading relay network in Fig. \ref{relay} with that of the Q-inverse soft-receiver when $m=1$. The generator matrices of the 2- and 3-user relay networks are given in (\ref{G2}) and (\ref{G3}), respectively. It is observed that the Q-inverse equivalent soft-receiver displays a very close error performance to the optimal soft-receiver.

In Figures 1-7, the simulation results regarding various receiver structures are presented. Next, various analytical results will be compared with simulation results in Figures 8-10.

Fig. \ref{2_3user_GF2_union} compares the analytical union bound BER performances of the 2- and 3-user GF(2) NC DetF minimum equivalent and Q-inverse equivalent relay networks with the optimum soft-receiver BER performances of the 2- and  3-user relay networks in a Nakagami-m fading environment when $m=1$. The 2- and  3-user relay networks' generator matrices are given in (\ref{G2}) and (\ref{G3}), respectively. The minimum and Q-inverse equivalent relay networks' union bounds are very close to the relay network's optimal performance.

Fig. \ref{2user_GF4_union} compares the analytical union bound SER performances of the 2-user GF(4) NC DetF minimum equivalent and Q-inverse equivalent relay networks with the optimum soft-receiver's SER performance in Nakagami-m fading with $m=1$. The generator matrix of the relay networks is given in (\ref{G1}). The union bound SER performances of the Q-inverse and minimum equivalent relay networks are very close to each other and approach the optimum soft-receiver SER performance at high SNR values.

Fig. \ref{GF2_Nakagami_union} compares the analytical union bound BER performance of the 2-user GF(2) NC DetF Q-inverse equivalent relay network with the optimum soft-receiver's BER performance in a Nakagami-m fading fading environment when $m=2$. The relay network's generator matrix is given in (\ref{G2}). The union bound BER performance of the Q-inverse equivalent relay network is very close to the optimum soft-receiver BER performance. Thus, the Q-inverse equivalent relay channel method provides a close approximation of the optimum receiver's error rate in a Nakagami-m fading environment.

\section{Conclusion}

This paper presents simple means of analyzing the error rate performance of a general wireless cooperative GF($q$) NC DetF relay network with known relay error statistics at the destination using equivalent channel models. The approximate error rate of the relay network is derived with the use of equivalent relay channel models, where the minimum and Q-inverse equivalent relay channel models are adapted to $N$-user GF($q$) NC DetF Nakagami-$m$ fading relay channels. The equivalent channel models are used in developing high-performance ML receivers for Nakagami-m fading channels, which are shown to exhibit very close error performances to the relay network's optimum receiver.

\bibliographystyle{IEEEtran}
\bibliography{Biblio}

\begin{thebibliography}{10}
\providecommand{\url}[1]{#1}
\csname url@rmstyle\endcsname
\providecommand{\newblock}{\relax}
\providecommand{\bibinfo}[2]{#2}
\providecommand\BIBentrySTDinterwordspacing{\spaceskip=0pt\relax}
\providecommand\BIBentryALTinterwordstretchfactor{4}
\providecommand\BIBentryALTinterwordspacing{\spaceskip=\fontdimen2\font plus
\BIBentryALTinterwordstretchfactor\fontdimen3\font minus
  \fontdimen4\font\relax}
\providecommand\BIBforeignlanguage[2]{{%
\expandafter\ifx\csname l@#1\endcsname\relax
\typeout{** WARNING: IEEEtran.bst: No hyphenation pattern has been}%
\typeout{** loaded for the language `#1'. Using the pattern for}%
\typeout{** the default language instead.}%
\else
\language=\csname l@#1\endcsname
\fi
#2}}

\bibitem{Proakis2000}
J.~G. Proakis, \emph{Digital Communications}, 4th~ed.\hskip 1em plus 0.5em
  minus 0.4em\relax McGraw-Hill, 2000.

\bibitem{Liu2008}
K.~J.~R. Liu, A.~K. Sadek, W.~Su, and A.~Kwasinski, \emph{Cooperative
  Communications and Networking}.\hskip 1em plus 0.5em minus 0.4em\relax
  Cambridge University Press, 2008.

\bibitem{Hasna2004}
M.~O. Hasna and M.-S. Alouini, ``{Harmonic Mean and End-to-End Performance of
  Transmission Systems with Relays},'' \emph{IEEE Transactions on
  Communications}, vol.~52, no.~1, 2004.

\bibitem{Wang2007}
T.~Wang, A.~Cano, G.~B. Giannakis, and J.~N. Laneman, ``{High-Performance
  Cooperative Demodulation With Decode-and-Forward Relays},'' \emph{IEEE
  Transactions on Communications}, vol.~55, no.~7, pp. 1427--1437, 2007.

\bibitem{Li2010}
Q.~You, Y.~Li, and Z.~Chen, ``{Joint Relay Selection and Network Coding Using
  Decode-and-Forward Protocol in Two-Way Relay Channels},'' in
  \emph{GLOBECOM'10}, 2010, pp. 1--6.

\bibitem{Benjillali2010}
M.~Benjillali and L.~Szczecinski, ``{Detect-and-Forward in Two-Hop Relay
  Channels: A Metrics-Based Analysis},'' \emph{IEEE Transactions on
  Communications}, vol.~58, no.~6, pp. 1729--1736, June 2010.

\bibitem{Chen2006}
J.~N. Chen, D.~Laneman, ``Modulation and demodulation for cooperative diversity
  in wireless systems,'' \emph{IEEE Transacations on Wireless Communications},
  vol.~5, no.~7, pp. 1785 -- 1794, July 2006.

\bibitem{Gao2011}
Y.~Gao, J.~Ge, and C.~Han, ``{Performance Analysis of Differential Modulation
  and Relay Selection with Detect-and-Forward Cooperative Relaying},''
  \emph{IEEE Communications Letters}, vol.~15, no.~3, pp. 323--325, 2011.

\bibitem{Yuan2010}
J.~Yuan, Y.~Li, and L.~Chu, ``{Differential Modulation and Relay Selection with
  Detect-and-Forward Cooperative Relaying},'' \emph{IEEE Transactions on
  Vehicular Technology}, vol.~59, no.~1, 2010.

\bibitem{Chu2008}
L.~Chu, Z.~Chen, and Y.~Li, ``{Performance Analysis and Code Design of
  Distributed Space-time Trellis Codes for the Detection-And-Forward
  Systems},'' in \emph{Australian Communications Theory Workshop}, February
  2008, pp. 144 -- 149.

\bibitem{Ahlswede2000}
R.~Ahlswede, N.~Cai, S.-y. R.~Li, and R.~W. Yeung, ``{Network Information
  Flow},'' \emph{IEEE Transactions on Information Theory}, vol.~46, no.~4, pp.
  1204--1216, 2000.

\bibitem{Lei}
Z.~Lei, Z.~Sihai, and Z.~Wuyang, ``{Cooperative Communication with Network
  Coding in GF(2n)},'' in \emph{IET International Communication Conference
  Wireless Mobile and Computing (ICCWMC 2009)}, 2009, pp. 586 -- 589.

\bibitem{Xiao2009b}
M.~Xiao, M.~Skoglund, ``M-user cooperative wireless communications based on
  nonbinary network codes,'' in \emph{IEEE Information Theory Workshop 2009
  (ITW '09)}, June 2009, pp. 316--320.

\bibitem{Xiao2009a}
------, ``Design of network codes for multiple-user multiple-relay wireless
  networks,'' in \emph{IEEE International Symposium on Information Theory 2009
  (ISIT 2009)}, 2009, pp. 2562 -- 2566.

\bibitem{Rebelatto2010}
J.~L. Rebelatto, B.~F. Uchoa~Filho, Y.~Li, and B.~Vucetic, ``Generalized
  distributed network coding based on nonbinary linear block codes for
  multi-user cooperative communications,'' in \emph{International Symposium on
  Information Theory 2010 (ISIT 2010)}, 2010, pp. 943--947.

\bibitem{Rebelatto2011}
J.~Rebelatto, B.~Uchoa-Filho, Y.~Li, and B.~Vucetic, ``Adaptive distributed
  network-channel coding,'' \emph{IEEE Transacations on Wireless
  Communications}, vol.~10, no.~9, pp. 2818 -- 2822, September 2011.

\bibitem{Xiao2009}
M.~Xiao and T.~Aulin, ``{Optimal Decoding and Performance Analysis of a Noisy
  Channel Network with Network Coding},'' \emph{IEEE Transactions on
  Communications}, vol.~57, no.~5, pp. 1402--1412, 2009.

\bibitem{Iezzi2011}
M.~Iezzi, M.~Di~Renzo, and F.~Graziosi, ``{Closed-Form Error Probability of
  Network-Coded Cooperative Wireless Networks with Channel-Aware Detectors},''
  in \emph{Global Telecommunications Conference 2011 (GLOBECOM 2011)}, IEEE,
  Ed., December 2011, pp. 1--6.

\bibitem{Iezzi2012}
------, ``Diversity, coding and multiplexing trade-off of network-coded
  wireless cooperative networks,'' in \emph{IEEE International Conference on
  Communication (ICC) 2012}, Ottawa, Canada, June 2012.

\bibitem{Kim2011}
D.~Kim, H.-M. Kim, and G.-H. Im, ``{Improved Network-Coded Cooperative
  Transmission with Low-Complexity Adaptation to Wireless Channels},''
  \emph{IEEE Transacations on Communications}, vol.~59, no.~10, pp. 2916--2927,
  October 2011.

\bibitem{Nasri2010}
A.~Nasri, R.~Schober, and M.~Uysal, ``Error rate performance of network-coded
  cooperative diversity systems,'' in \emph{IEEE Global Telecommunications
  Conference 2010 (GLOBECOM 2010)}, December 2010, pp. 1--6.

\bibitem{Safak2013}
I.~{\c{S}}afak, ``{Equivalent Relay Channel Approach to GF(q) Network Coded
  Wireless Relay Networks},'' Ph.D. dissertation, Hacettepe University, January
  2013.

\bibitem{Papoulis2002}
S.~U. Papoulis, A.~Pillai, \emph{Probability, Random Variables and Stochastic
  Processes}, 4th~ed.\hskip 1em plus 0.5em minus 0.4em\relax McGraw-Hill, 2002.

\bibitem{Simon2000}
M.-S. Simon, M. K.~Alouini, \emph{Digital Communication over Fading Channels: A
  Unified Approach to Performance Analysis}.\hskip 1em plus 0.5em minus
  0.4em\relax John Wiley \& Sons, Inc., 2000.

\bibitem{Goldsmith2005}
A.~Goldsmith, \emph{Wireless Communications}.\hskip 1em plus 0.5em minus
  0.4em\relax Cambridge University Press, 2005.

\bibitem{Gradshteyn2000}
L.~Gradshteyn and I.~Ryzhik, \emph{Table of Integrals, Series and Products},
  6th~ed.\hskip 1em plus 0.5em minus 0.4em\relax Academic Press Inc., 2000.

\end{thebibliography}

\end{document}